# Identification of multidisciplinary research based upon dissimilarity analysis of journals included in reference lists of Wageningen University & Research articles

Veller, M.G.P. van

# Identification of multidisciplinary research based upon dissimilarity analysis of journals included in reference lists of Wageningen University & Research articles

Marco G.P. van Veller[1]

[1]Wageningen University & Research - Library, Droevendaalsesteeg 2, 6708 PB, Wageningen, the Netherlands

## Abstract

This paper discusses the identification of journal articles that probably report on multidisciplinary research. Identification of these articles may be important for strategic purposes for the institution where the research is performed or for the evaluation of researchers or groups.

In order to identify possibly multidisciplinary research, this paper describes an analysis on the journals from which articles have been cited in the journal articles published by Wageningen University & Research (WUR) staff in 2006-2015. The journals with cited articles are inventoried from the reference lists of the WUR journal articles. For each WUR article a mean dissimilarity is calculated between the journal in which it has been published and the journals that contain the cited articles. Dissimilarities are derived from a large matrix with similarity values between journals, calculated from co-citations to these journals from the same WUR articles published in 2006-2015.

For 21,191 WUR articles published in 2,535 journals mean dissimilarities have been calculated. For WUR articles with high mean dissimilarities this paper shows that they often are published in multidisciplinary journals. WUR articles with high mean dissimilarities also are found in non-multidisciplinary (research field specific) journals. For these articles (with high mean dissimilarities) this paper shows that citations are often made to more various research fields than for articles with lower mean dissimilarities.

The mean dissimilarities, calculated per WUR article, also can be aggregated for the journals in which they have been published. This results in a listing of journals than can be ordered on the mean dissimilarity of the WUR articles that have been published in them. The analysis described in this paper shows that journals for which a high mean dissimilarity is calculated tend to have a multidisciplinary scope.

## Keywords

## Introduction

Wageningen University & Research (WUR) is an international research and knowledge centre in the Netherlands that consists of a university and various research institutes. The mission of WUR is to explore the potential of nature to improve the quality of life. To accomplish this, various research activities are undertaken by staff and students around three related core areas (WUR, 2016):

- food, feed and biobased production;
- natural resources and living environment; and
- society and well-being.

At WUR work 6,500 staff and study 10,000 students from over 100 countries. These staff and students present their research in publications of various kinds (e.g. reports, books, book chapters, conference proceedings, journal articles, etc.). Once registered in the Research Information System (RIS) of WUR, these publications are included in WUR's institutional repository (called Staff publications). Of the publications produced by WUR staff and students (i.e. WUR authors) about one third consists of (refereed) articles in scholarly journals (Staff publications, 2017).

To get an idea of the research topics that are investigated at WUR, 3,020 articles that have been published in 2015 by WUR authors have been looked up in Web of Science. From the bibliographic data (title and abstract) of these articles a term map based on text data is made with VOSviewer (Eck & Waltman, 2016). Figure 1 shows this term map.

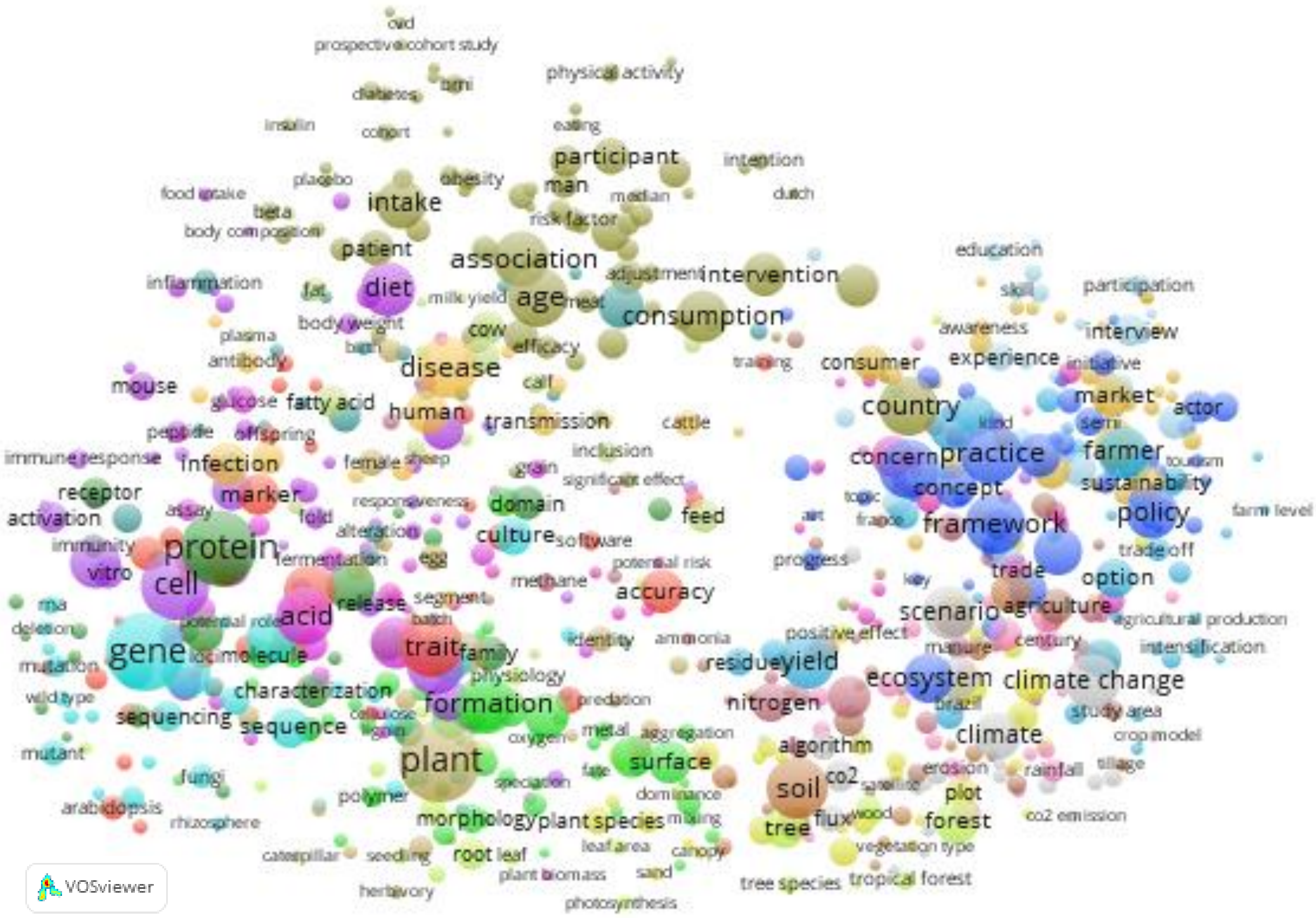

Figure 1: Term map for 3,020 WUR articles indexed in Web of Science and published in 2015.

The term map shows that (e.g.) gene, protein, plant, disease, ecosystem, farmer and climate change are important terms that WUR authors used in their articles. These terms represent a selection of the most important topics that have been investigated at WUR in 2015. Further, the map in figure 1 shows 24 clusters of term maps. Depending on the settings of VOsviewer while calculating the term map, the number of clusters may vary. Further examination of some of the clusters show that they each indicate a particular research field that is examined at WUR (e.g. the red cluster corresponds with breeding and genetics studies and the grey cluster corresponds with climate studies).

As indicated by the map in figure 1, the articles that are published by WUR authors give information on the topics and research field that are investigated at WUR. For some of the articles it is clear which research field they belong to because they are published in journals that typically represent scientific output for a particular research field. For other articles, however, it is more difficult to assign them to a particular research field because they are published journals that have a multidisciplinary scope. Also, it can be possible that, although published in a field-specific journal, articles may deal with topics from various research fields and describe research that has a multidisciplinary setup.

For a complete exploration of themes investigated in one or more of the core areas listed above, often a multidisciplinary research approach is necessary. To identify multidisciplinary research this paper proposes a method to select WUR articles that are dissimilar to other WUR articles in terms of including journals in their references (i.e. journal that include the articles that have been cited in the WUR articles) that are not very similar to the journal where the articles have been published in. The references of the WUR articles include information on the journals that have been important for the authors for the research they describe in their articles. Therefore, analysis of the journals included in the references provides information on the nature of the research fields that are dealt with in the articles.

After selection of dissimilar WUR articles this paper examines whether these articles describe multidisciplinary research. Also, with the calculation of the dissimilarities of articles, this paper describes how a list of scholarly journals can be made that on average contain higher numbers of dissimilar WUR articles. Some of these journals have a multidisciplinary scope (for example PlosOne). Other journals, however, represent a particular research field but mostly contain WUR articles that are quite dissimilar and that may represent multidisciplinary research.

## Materials and methods

In an earlier paper, Veller (2013) describes a methodology to get customized lists of journals that have been cited in WUR articles over a certain period of time. In another paper Veller & Gerritsma (2015) make an analysis of the number of times journals have been co-cited in articles published by WUR authors in 2006-2013. Based upon this analysis they propose a measure of similarity between citing journals. The similarity measure that Veller & Gerritsma (2015) propose is based upon relative abundances of cited journals in the references of the articles published in the (citing) journals for which the similarity is calculated. The more the relative abundances of co-cited journals in the references of articles published in two citing journals are, the higher the similarity between these citing journals is.

This paper applies the similarity measure proposed by Veller and Gerritsma (2015), but in a modified way. Where in Veller & Gerritsma (2015) the similarities are calculated for journals with WUR articles (i.e. the citing journals), this paper calculates similarities between the journals included in the references of the WUR articles (i.e. the cited journals). To calculate these similarities a list of cited journals is made. For each of these cited journals it is inventoried how many times it has been cited in the WUR articles. This inventory is made per combination of a cited and a citing journal, resulting in a list with the number of times a cited journal has been cited by articles published in a citing journal.

From the list similarities between cited journals are calculated based upon shares of citations from articles published in the citing journals. There more two journals are cited by comparable shares of citations from articles published in the same citing journals, the higher the similarity between these cited journals is. The calculation of the similarity ($S_{gh}$) between cited journal g and cited journal h is formalized as follows.

$$S_{gh} = 1 - \frac{\sum_{j=1}^{n_j} \left| \frac{c_{jg}}{\sum_{j=1}^{n_j} c_{jg}} - \frac{c_{jh}}{\sum_{j=1}^{n_j} c_{jh}} \right|}{2}$$

$S_{gh}$=similarity between cited journal g and cited journal h

$c_{jg}$= number of citations to journal g in articles published in citing journal j .

$n_j$=number of citing journals

$c_{jh}$= number of citations to journal h in articles published in citing journal j.

The calculated similarities between n cited journals are collected in a symmetric $n_g$ x $n_h$ matrix. At the diagonal the value 1 represents complete similarity of these journals with themselves. In this matrix the similarity $S_{gh}$ between two journals g and h is listed in the cell corresponding with row g and column h.

The similarities between the journals can be visualized with VOSviewer (Eck & Waltman, 2016). Only journals with minimal 10 citations from WUR articles published in 2006-2015 are included in the visualization. For the cluster analysis and mapping the following parameter settings are applied:

- Mapping attraction: 2
- Mapping repulsion: 1
- Clustering resolution: 1.10
- Minimum cluster size: 10
- Normalization method 1

For all references in WUR articles published in 2006-2015, similarities are obtained from the symmetric $n_g$ x $n_h$ matrix. Hereby, for each reference, the journal in which the article has been published corresponds with row g and the journal to which is cited corresponds with column h. Per WUR article, the mean is calculated over all similarities obtained for its references. When a particular journal has been included in the references of an article more than once, for each occurrence the similarity of this journal with the citing journal is included in the calculation of the mean. The mean similarity of all references in a WUR article is transformed into a dissimilarity value by subtracting it from one. The calculation of the mean dissimilarity ($\overline{D}_{gi}$) over references j for article i published in journal g is formalized as follows.

$$\overline{D}_{gi} = 1 - \frac{\sum_{j=1}^{n_j} S_{gh_j}}{n_j}$$

$\overline{D}_{gi}$=mean dissimilarity of article i published in journal g

$S_{gh_j}$=similarity between journal g (in which the citing article i has been published) and journal h (in which the cited article has been published that is listed in reference j of article i)

$n_j$=number of references for article i published in journal g

The mean dissimilarities for WUR articles published in the same journal g can be used to calculate a mean dissimilarity for journal g ($\overline{D}_g$). This is formalized as follows.

$$\overline{D}_g = \frac{\sum_{i=1}^{n_i} \overline{D}_{gi}}{n_i}$$

$\overline{D}_g$=mean dissimilarity of journal g

$n_i$=number of articles published in journal g

## Results

In the period 2006-2015 WUR authors published 21,191 articles in 2,535 ISI journals (i.e. journals covered by Web of Science). From these WUR articles in total 925,809 citations to 13,257 journals have been collected. For each of these cited journals it is inventoried how many times it has been cited from each of the journals in which the WUR articles were published. Based upon having received citations from the same (citing) journals a similarity value ($S_{gh}$) was calculated with R (R Core Team, 2015) for each pairwise combination of cited journals g and h according to the formula listed above. The similarities were saved in a similarity matrix of 13,257 x 13,257 cited journals. The R-script for the calculation of this similarity matrix is listed in appendix 1.

Figure 2 shows a network that represents the similarity between the cited journals. For this network only journals are selected that received minimal 10 citations from the WUR articles in 2006-2015. This selection results in 5,015 of the 13,257 cited journals. Together these more than five thousand journals received 97% of all citations from the WUR articles.

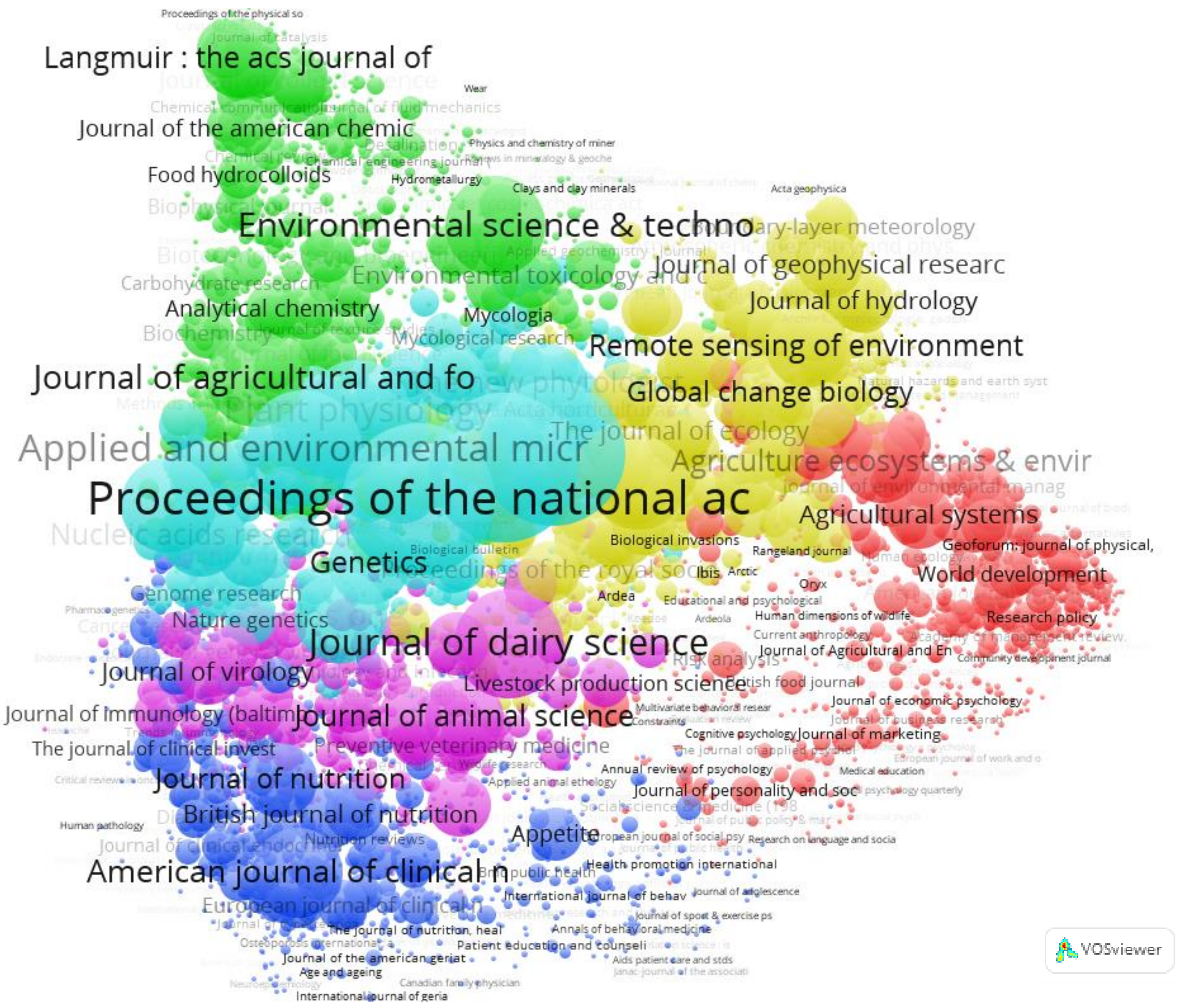

Figure 2: Network of 5,015 journals based upon a similarity analysis on citations to these journals from research articles published by WUR in 2006-2015.

Figure 2 shows that the cited journals fall in six different clusters. Further investigation of these clusters reveals that each broadly corresponds with main scientific disciplines which are investigated at WUR:

- the red cluster corresponds with journals in agroecosystem sciences and agronomy;
- the green cluster corresponds with journals in chemical sciences and biotechnology;
- the yellow cluster corresponds with journals in environmental sciences;
- the light blue cluster corresponds with journals in plant sciences;
- the dark blue cluster corresponds with journals in health and nutrition; and
- the purple cluster corresponds with journals in animal sciences.

By adjusting the parameter settings for mapping in VOSviewer it is possible to distinguish more clusters that each comprise a smaller set of journals corresponding with a certain scientific discipline.

From the more than five thousand journals that are presented in the network in figure 2 only the journals that contain more than two WUR articles published in 2006-2015 are selected. This selection includes 1,175 journals with 20,362 WUR articles (96% of the articles that are analyzed in this paper).

Based upon the position of each journal in the network presented in figure 2 a sub-network is presented in figure 3 with the total number of WUR articles per journal proportional to the size of the bubbles.

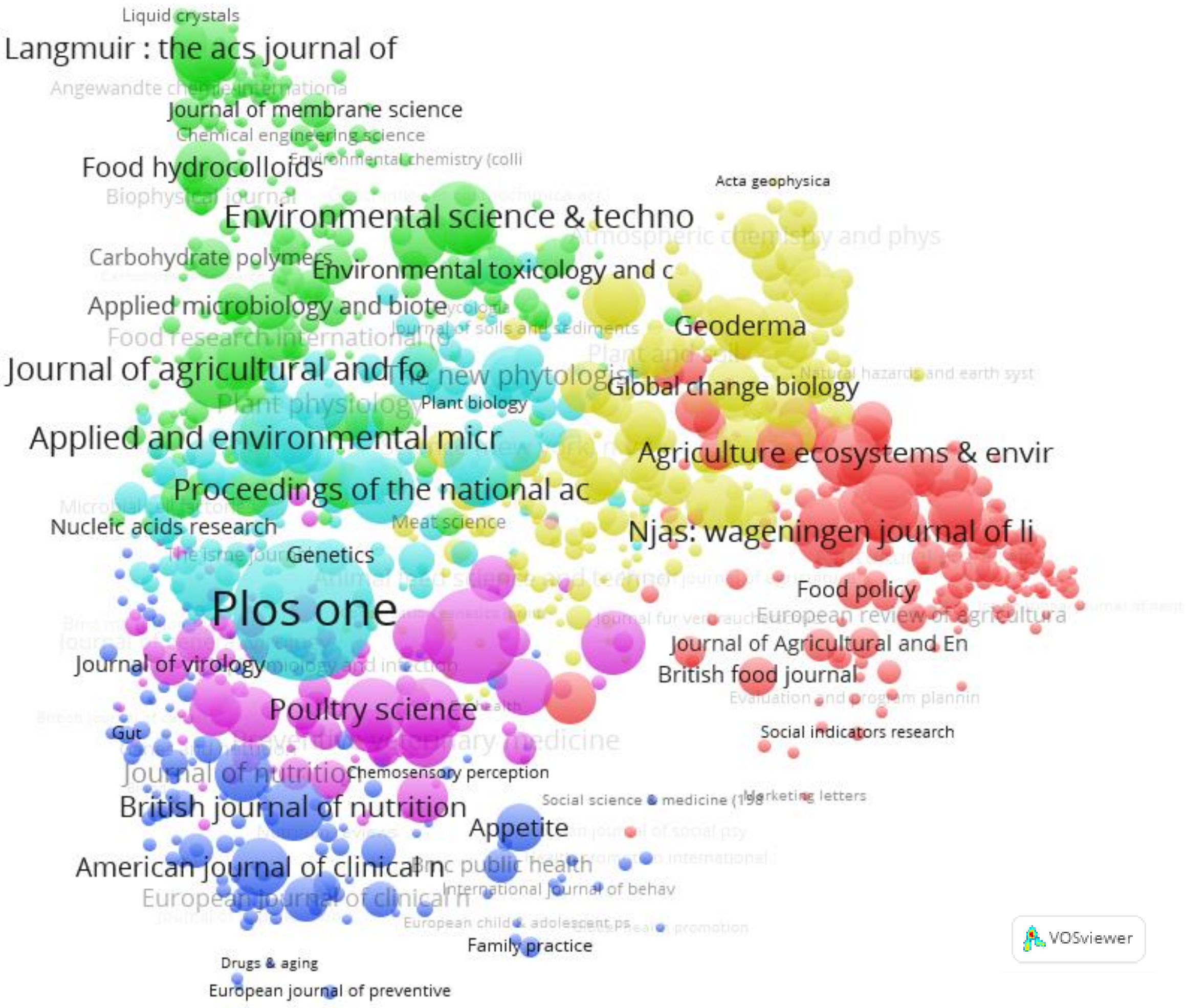

Figure 3: Network of 1,175 journals based upon a similarity analysis on citations to these journals from research articles published by WUR in 2006-2015 with mapping of the number of articles published per journal (score proportional to the size of a bubble).

From figure 3 it follows that the top 20 of journals with the highest number of WUR articles published in 2006-2015 can be found back in all six different clusters. Table 1 lists these 20 journals with information on the number of WUR articles and the cluster they are placed in based upon the similarity analysis described above.

Table 1: Top 20 of journals with the largest number of WUR articles published in 2006-2015 with information the number of articles, clustering and mean dissimilarity.

| Journal | Number of WUR articles in 2006-2015 | Clustering according to network presented in figure 2 | Mean dissimilarity with journals cited from articles |
|---|---|---|---|
| Plos one | 611 | plant sciences (light blue) | 0.76 |
| Journal of dairy science | 282 | animal sciences (purple) | 0.45 |
| Journal of agricultural and food chemistry | 173 | chemical sciences and biotechnology (green) | 0.64 |
| Environmental science & technology | 163 | chemical sciences and biotechnology (green) | 0.83 |
| Applied and environmental microbiology | 160 | plant sciences (light blue) | 0.56 |
| NJAS: Wageningen journal of life sciences | 154 | agroecosystem sciences and agronomy (red) | 0.81 |
| Agricultural systems | 148 | agroecosystem sciences and agronomy (red) | 0.61 |
| Langmuir : the acs journal of surfaces and colloids | 144 | chemical sciences and biotechnology (green) | 0.53 |
| Proceedings of the National Academy of Science of the United States | 140 | plant sciences (light blue) | 0.85 |
| Poultry science | 136 | animal sciences (purple) | 0.49 |
| Journal of animal science | 128 | animal sciences (purple) | 0.55 |
| Livestock science | 127 | animal sciences (purple) | 0.64 |
| British journal of nutrition | 124 | health and nutrition (dark blue) | 0.63 |
| Bmc genomics | 121 | plant sciences (light blue) | 0.65 |
| Agriculture ecosystems & environment | 120 | agroecosystem sciences and agronomy (red) | 0.89 |
| American journal of clinical nutrition | 107 | health and nutrition (dark blue) | 0.55 |
| Geoderma | 105 | environmental sciences (yellow | 0.57 |
| The new phytologist | 105 | plant sciences (light blue) | 0.85 |
| Ices journal of marine science | 103 | environmental sciences (yellow) | 0.47 |
| Animal : an international journal of animal bioscience | 102 | animal sciences (purple) | 0.61 |

In order to calculate the mean dissimilarity ($\overline{D}_{gi}$) over all references for article i published in journal g the similarity matrix of 13,257 x 13,257 cited journals is used. Hereby, for each of the 21,191 WUR articles firstly similarity values between the journal in which the article has been published and each of the journals that contains the cited article (included in the references) are obtained from the matrix and averaged over all cited articles in the WUR article. Secondly, the mean similarity value is transformed to an mean dissimilarity by subtraction from 1.The R-script to calculate the mean dissimilarity is listed in appendix 2.

Figure 4 shows the distribution of the mean dissimilarity per WUR article.

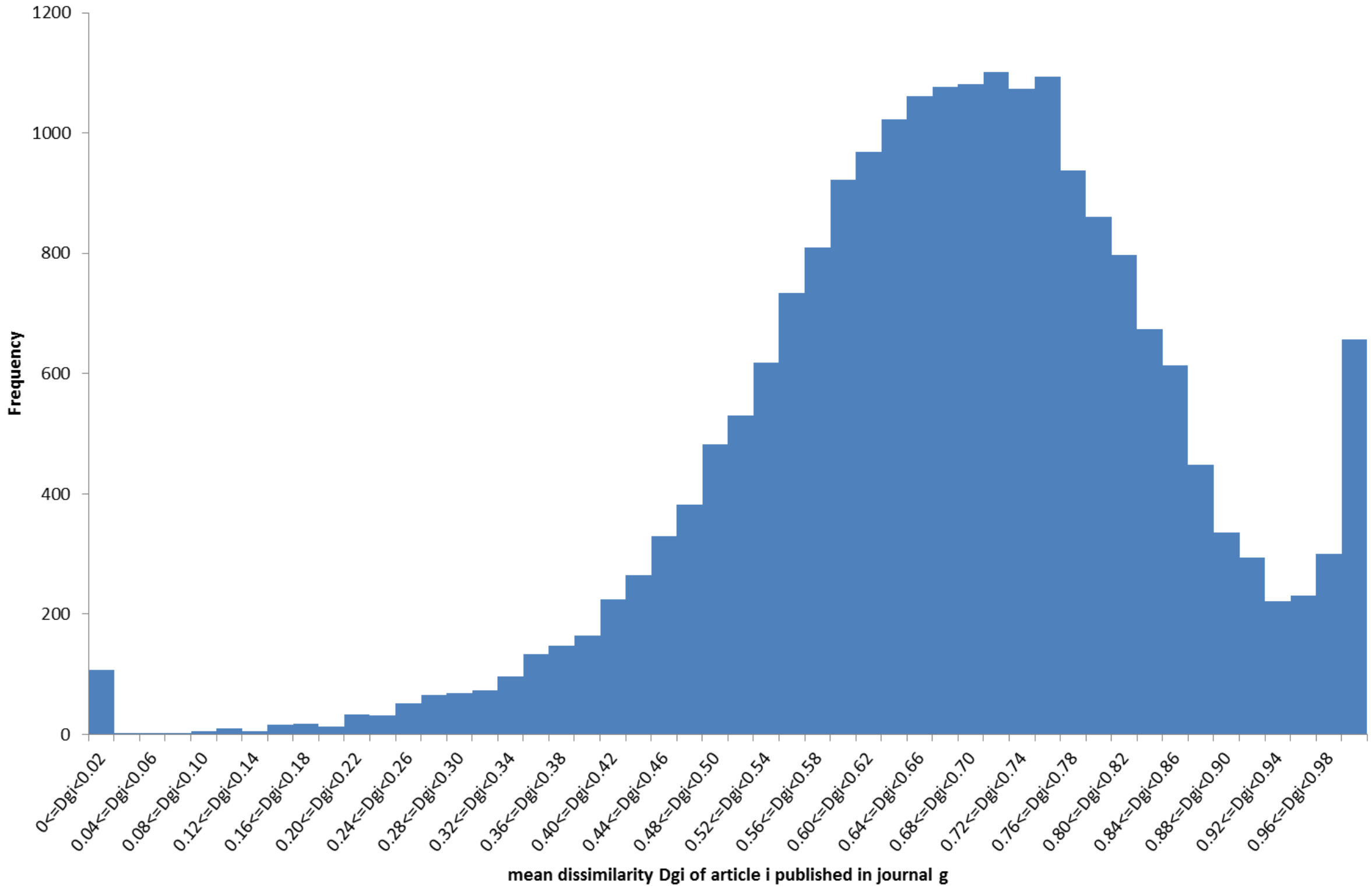

Figure 4: Distribution of the mean dissimilarity of 21,191 WUR articles calculated from dissimilarities between the journal in which each article is published and the journals from which articles have been cited in each WUR article.

The histogram in figure 4 shows that the mean dissimilarity per WUR article lies between 0 and 1. The most frequent mean dissimilarity lies around 0.70. For a considerable number of WUR articles the mean dissimilarity is either very low (107 articles with values below 0.02) or high (657 articles with values above 0.98). For the articles with a low mean dissimilarity further inspection of the data reveals that these articles have only one or a few references. The articles with a high dissimilarity have equal amounts of references as the other WUR articles but seem to be published in multidisciplinary journals (e.g. Proceedings of the national academy of sciences of the united states, Environmental research letters, Plant science or Current biology).

The mean dissimilarity ($\overline{D}_{gi}$) per article i is averaged over all articles published in journal g resulting in the mean dissimilarity ($\overline{D}_{g}$) for journal g. This mean dissimilarity per journal is mapped as score value in the network that represents similarities between journals with more than two WUR articles published in 2006-2015 (see figure 3). Figure 5 shows this mapping.

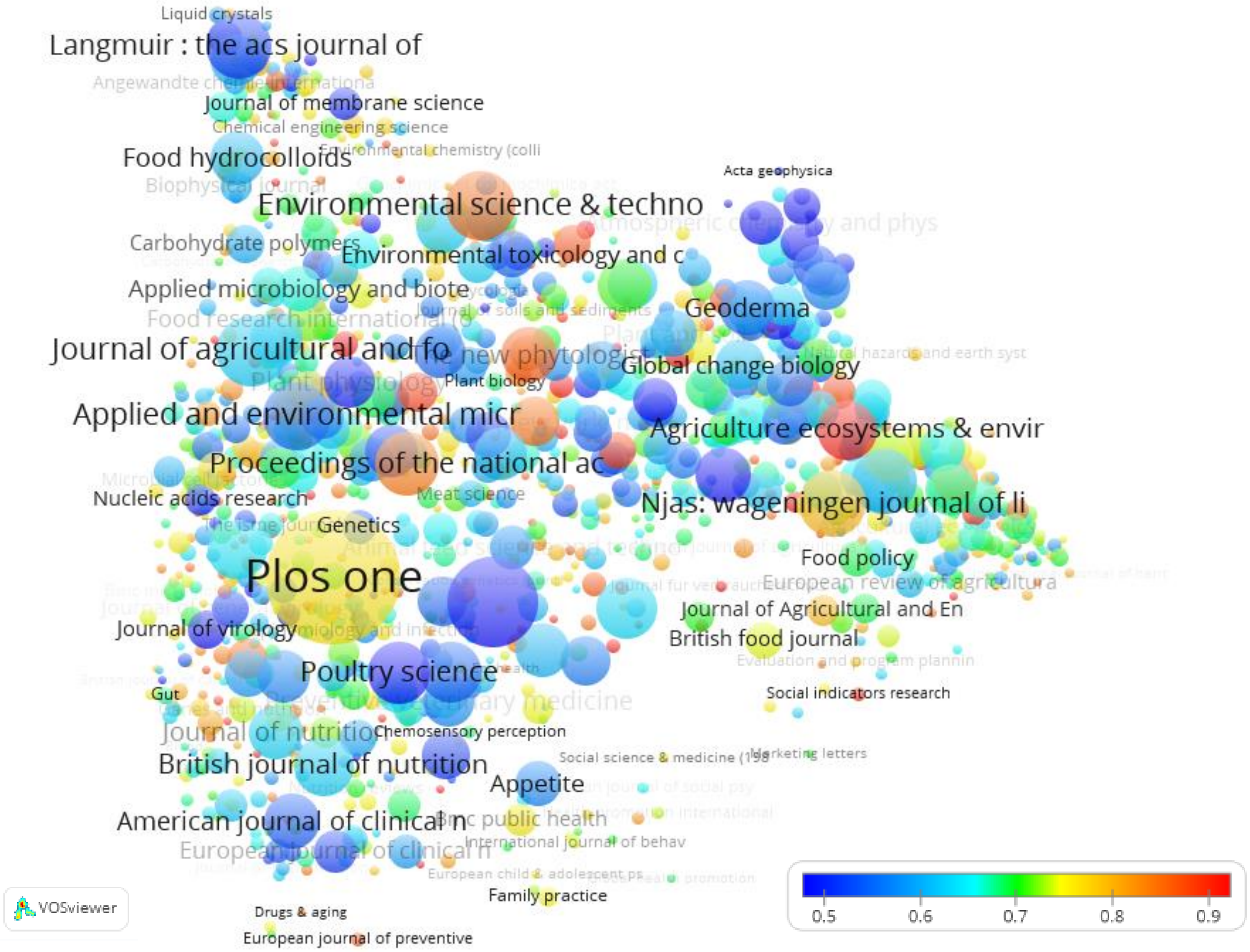

Figure 5: Network of 1,175 journals that contain more than two WUR articles published in 2006-2015 with mapping of the mean dissimilarity per journal (score corresponds with the colour of a bubble) and the number of articles published per journal (score proportional to the size of a bubble).

The network in figure 5 visualizes the relationships between the selected 1,175 journals based upon similarity in having been cited by the same journals in the WUR articles. The more two journals are placed together, the more similar they are. The size of the bubbles in figure 5 is proportional to the size of the article output in each of the journals by WUR authors in 2006-2015. The red journals in figure 5 have high values for the mean dissimilarity per journal.

Journals with high values for the mean dissimilarity per journal (more than 0.75) and a considerable number of WUR articles (more than 100 published between 2006 and 2015) are Agriculture ecosystems & environment, The new phytologist, Environmental science & technology, Proceedings of the national academy of sciences of the united states, Njas: wageningen journal of life sciences and Plos One. The mean dissimilarity per journal for the Top 20 of journals with the largest number of WUR articles published in 2006-2015 is included in the last column of table 1.

## Discussion and conclusions

The methodology and analysis described in this paper on WUR articles published in 2006-2015 shows that it is possible to calculate a mean dissimilarity per article based upon dissimilarities between the journal in which this article has been published and the journals that are listed in its references.

The distribution of the mean dissimilarity per article shows there are 107 WUR articles with mean dissimilarities below 0.02. Further inspection of these articles reveals they each have a very short reference list (on average three references) in which only articles are included that are published in the same journal. Contrary to these WUR articles with very low mean dissimilarities there are 657 articles with a mean dissimilarity above 0.98. When these articles with high dissimilarities are further examined, it is found that their reference lists on average contain 39 references per article. This is close to the average number of references per article (38) for all WUR articles that have been analyzed in this paper. Of the 657 articles with high dissimilarities there are 19 WUR articles with a dissimilarity of one. Further inspection of these articles reveals that they each have a short reference list (on average seven references) in which only articles are included that are published in other journals than the journal in which the WUR article is published.

The results described above show that the calculation of the mean dissimilarity per article is sensitive for the number of cited articles included in the reference list. For articles with short reference lists extreme values of dissimilarities can be found because the calculation of these dissimilarities is based upon only a few cited articles that all are either published in the same journal or different journals as the journal in which the citing article has been published. In order to correct for this anomaly, a selection of 19,789 WUR articles that each contain at least 10 references is made. These selected articles together contain 99% of all citations to journal articles that are inventoried from the WUR articles that are published in 2006-2015 and analyzed in this paper.

To investigate whether high mean dissimilarities for articles co-occur with publication of these articles in multidisciplinary journals, the selected articles have been divided in 10 classes based upon the deciles of their dissimilarities. Per class, the percentage of articles published in multidisciplinary journals have been calculated. For the identification of multidisciplinary journals, the category “Multidisplinary sciences” with 63 journals from the Journal Citations Reports have been used. Figure 6 shows the percentages of selected articles published in multidisciplinary journals per class (decile) of dissimilarities.

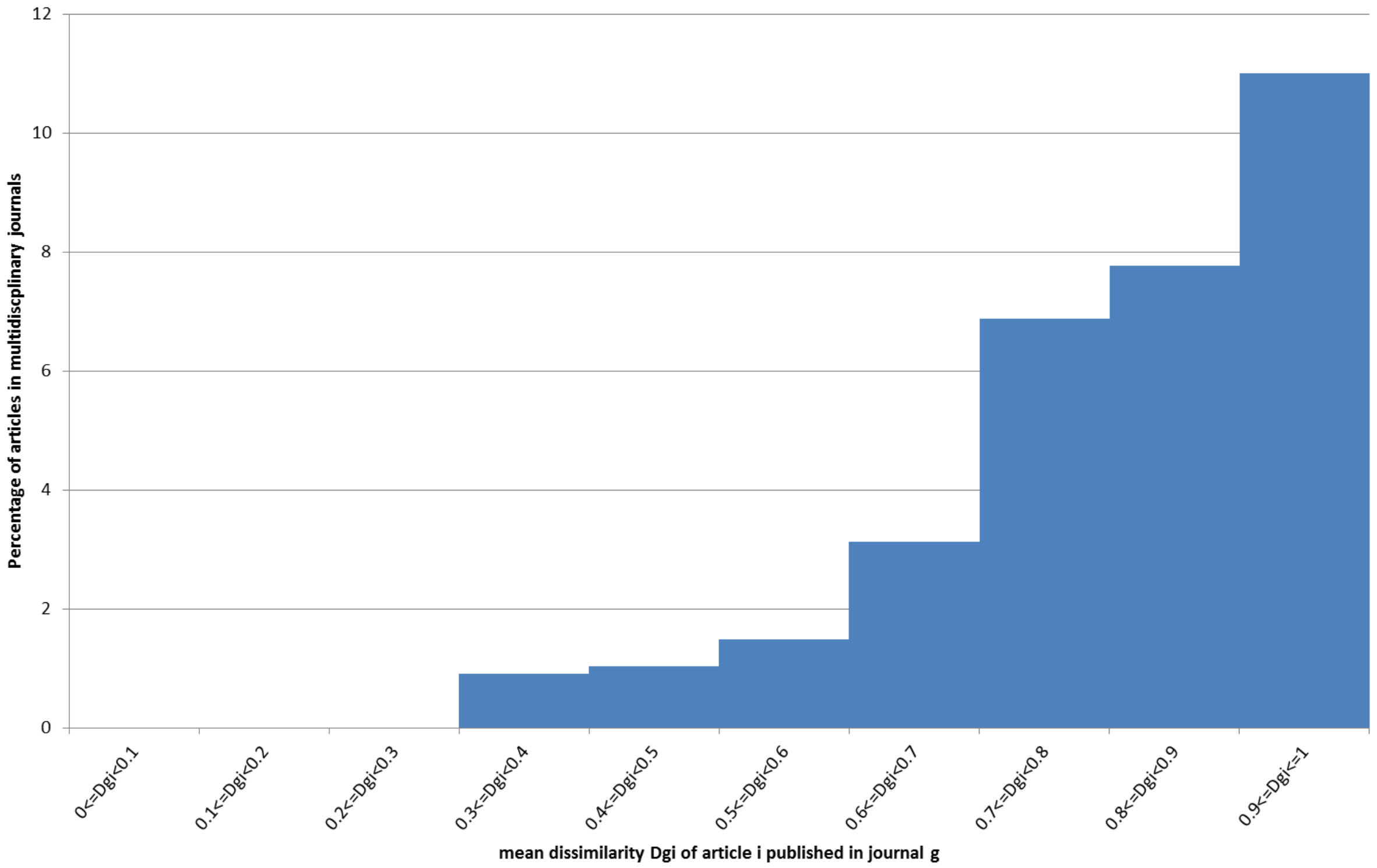

Figure 6: Percentages of WUR articles (each with at least 10 cited articles) published in multidisciplinary journals for 10 classes of articles based upon deciles of the mean dissimilarity of each article.

From figure 6 it follows that the classes with articles that have low dissimilarities also have low percentages of articles published in multidisciplinary journals. For the class of articles with the highest dissimilarity (above 0.90), 11% of the articles are published in multidisciplinary journals. These findings indicate that high values of dissimilarities for articles can be an indication that these articles have been published in multidisciplinary journals.

However, it may also be that an article describes multidisciplinary research but is published in a journal that only represents a particular research field. To illustrate this, two articles from the journal Ecology (ISSN: 0012-9658) are selected and represented in figure 7. In the Journal Citation Reports, the journal Ecology is classified in the category “Ecology”. Also, the website of the journal (Journal Ecology, 2017) indicates that this journal publishes articles that present ecological research and ecological phenomena in particular.

From the journal Ecology two WUR articles are selected; one with a low dissimilarity (0.38) and the other with a high dissimilarity (0.99). For both articles it is inventoried how many articles from which journals have been cited and to which categories in the Journal Citation Reports these journals (with cited articles) belong. Figure 7 shows the shares and variety in the categories (research fields) to which these citations have been made from each article.

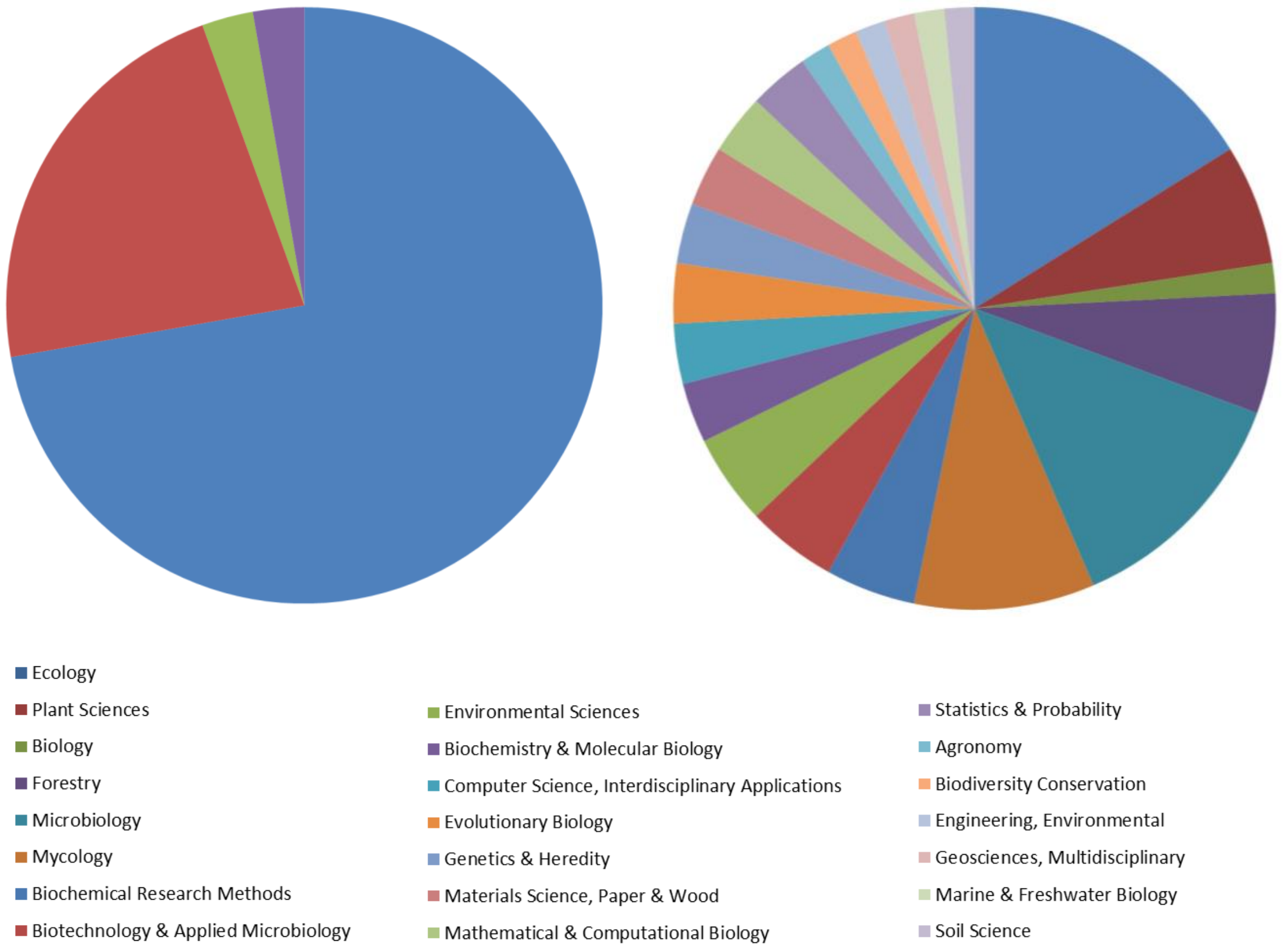

Figure 7: Shares of the categories (research fields) to which the journals (according to the Journal Citation Reports) belong that contain the articles cited in a WUR article with a low dissimilarity (left) and a WUR article with a high dissimilarity (right). Both WUR articles have been published in the journal Ecology and respectively cite 31 and 35 articles in journals for which category information can be obtained from the Journal Citation Reports.

From figure 7 it follows that although published in the same journals, the article with the high dissimilarity cites articles from more different categories  or research fields (22 categories for the article with the high dissimilarity as opposed to four categories for the article with the low dissimilarity). For the article with the low dissimilarity 72% of the citations are made to articles in the same category (research field) as the category (research field) in which the article has been published (i.e. Ecology). For the article with the high dissimilarity 72% of the citations are made in 11 categories, thereby covering a larger variety of research fields.

It is also investigated whether in general from articles with higher dissimilarities more different categories (research fields) are represented by the journals from which articles have been cited by WUR authors. This inventory of categories have been made on 18,853 WUR articles published in non-multidisciplinary journals with each at least 10 citations to journal articles. Figure 8 shows the mean number of different categories to which the journals with the cited articles belong for the 10 deciles (that also have been used in figure 6) for the mean dissimilarities of WUR articles.

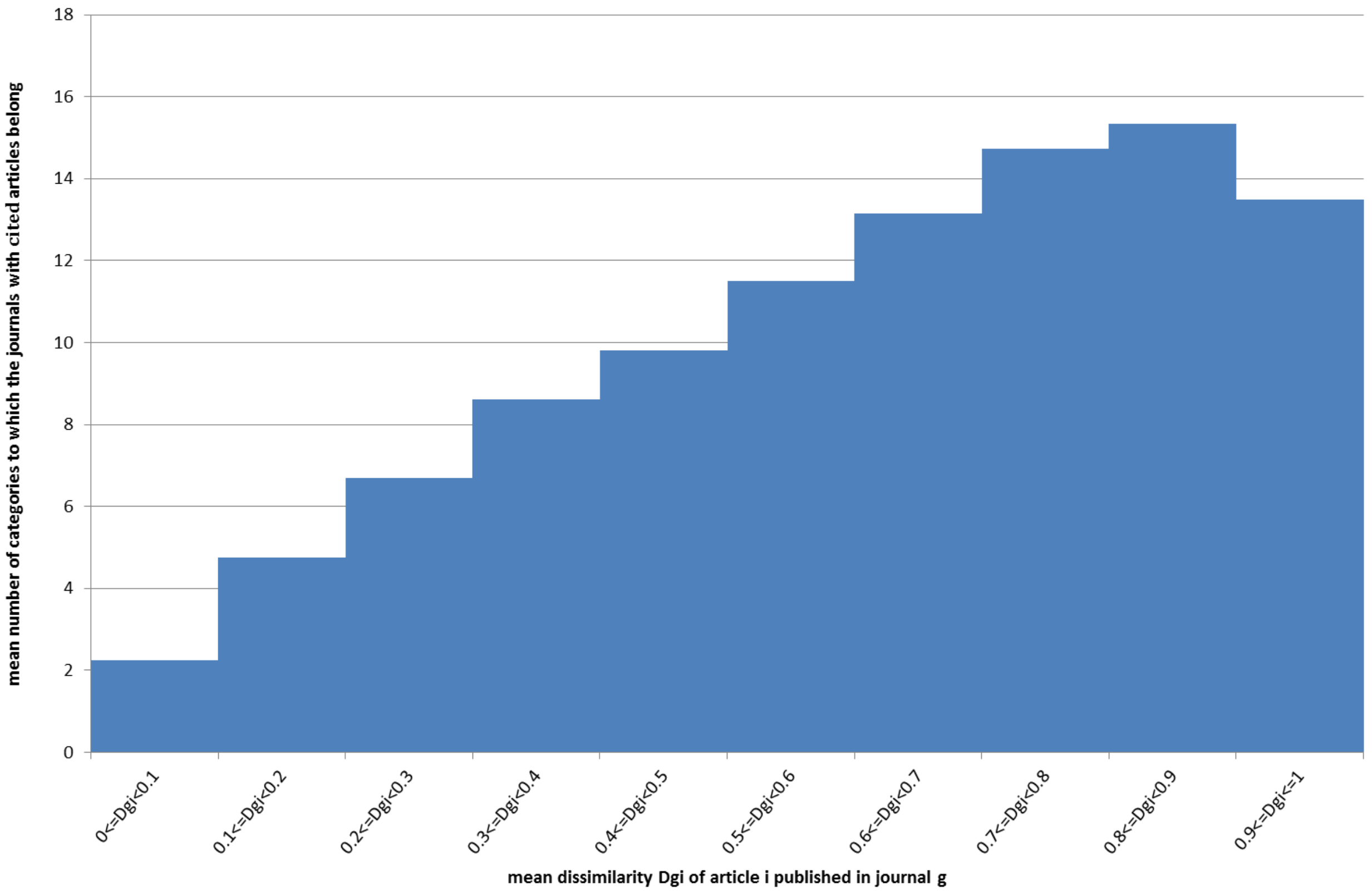

Figure 8: Mean number of categories (research fields) to which the journals with cited articles belong for 10 classes of WUR articles from which the citations have been made. The WUR articles have been published in non-multidisciplinary journals and each cite at least 10 journals articles. The classes are based upon deciles of the mean dissimilarities of the WUR articles.

In accordance with the findings for the two WUR articles published in the journal Ecology, figure 8 also shows that articles in journals that belong to more different categories (research fields) are cited from WUR articles with higher dissimilarities. These findings support that for WUR articles with high dissimilarities the indications are stronger that they report on multidisciplinary research, either described by publication in a multidisciplinary journal (see figure 6) or by publication of a multidisciplinary research in a non-multidisciplinary journal (see figures 7 and 8).

Based upon the methodology and analysis described in this paper one can identify WUR articles that possibly report on the findings of multidisciplinary research. Once the articles that possibly represent multidisciplinary research have been identified, the Research Information System of the research institution can be used to identify authors and research groups that have published these articles. By aggregation of articles for authors or research groups and selection of sets of articles with high values of mean dissimilarities it is possible to identify authors or research groups that often use a multidisciplinary approach in their research. This can be interesting for evaluation or research strategy purposes at the research institution.

The results described in this paper show that the dissimilarities that are calculated for articles can also be aggregated to calculate mean dissimilarities for journals. The mean dissimilarity per journal can be used to identify journals that have relatively larger shares of published articles that describe multidisciplinary research. The networks presented in figures 3 and 5 show that for WUR the journals with higher shares of “multidisciplinary articles” especially can be found in the scientific disciplines of chemical sciences and biotechnology, plant sciences and agroecosystem sciences and agronomy. It may be interesting to identify journals that are important to authors for publishing multidisciplinary research findings in particular fields that are important for WUR.

## Acknowledgements

The author thanks wishes to thank Ek'abo Omwana for inspiration on the writing of this paper.

## References

Eck, N.J. van & Waltman, L. (2016). VOSviewer version 1.6.5. Center for Science and Technology Studies. Leiden University. Available at: http://www.vosviewer.com/.

Journal Ecology (2017). Access date 13.04.2017. Available at: http://eu.wiley.com/WileyCDA/WileyTitle/productCd-ECY.html

R Core Team (2015). R: A language and environment for statistical computing. R Foundation for statistical Computing, Vienna, Austria. Available at: http://www.R-project.org/.

Staff publications (2017). Access date 13.04.2017. Available at : http://library.wur.nl/WebQuery/wurpubs/show.

Veller, M.G.P. van (2013). Analysis of journal usage by WUR staff members via article references. *Qualitative and quantitative methods in libraries*, 2: 231-244.

Veller, M.G.P. van & Gerritsma, W. (2015). Development of a journal recommendation tool based upon co-citation analysis of journals cited in WUR research articles. Qualitative and quantitative methods in libraries, 4: 233-257.

WUR (2016). Annual *report WUR 2015*. WUR. Wageningen. 122 pp.

## Appendix 1: R-script for calculation of the similarity matrix

Below, the R-script is given that has been used for calculating the similarity matrix for 13,257 journals based upon citations made to articles in these journals from 2,535 journals in which WUR authors published their research articles. The similarity matrix is used as input to VOSviewer to create a network for a selection of 5,015 cited journals(with each minimal 10 citations received from the WUR articles). Also, the matrix is used to obtain similarity values between journals that are used to calculate a mean dissimilarity per WUR article (see R-script in appendix 2). Commentary lines in this script start with “#”.

```
#Import datafile with three rows. The first column consists of journals
#(identified by 13257 unique numbers) that were cited in articles
published #by WUR in 2006-2015. The second consists of journals (via
2535 unique #numbers) in which WUR published and made the citations
from. The data #represent the number of references per journal. The
third column consists #of relative shares that have been brought to
each cited journal per #journal from which the citations were made.
Similarity is calculated for #the journals that were cited from the WUR
publications. Similarity is #based upon a modification on the Jaccard
index. Only select the data and #copy them to the clipboard. The
clipboard contents is copied to the #dataframe journals. You will have
to type in the statement below in order #to keep the data in the
clipboard for import.

journals<-read.table("clipboard")

JacSim<-matrix(,13257,13257)

m<-matrix(0,2535,2)

#Calculate similarities and place them in the matrix JacSim. Calculate
the #similarities between journals that were cited by WUR by
alfabetically #ordering these journals and replacing their names by a
number.This script #has been developed for 13257 cited journals from
WUR pubs. The journals #are compared for co-citations from journals
with WUR publications. The #list of cited journals in this script is
number 1 to 2535 after having #ordered them alphabetically.

for (k in 1:13257) {

print (k)

m[,]<-0

a<-data.frame(journals[2][journals[1]==k],journals[3][journals[1]==k])

a<-aggregate(a[2], by=a[1], sum)

m[a[,1],1]<-a[,2]

for (l in 1:k) {

print (l)
```

```
if (l==k) {

JacSim[k,l]<-1

}

else

{

b<-data.frame(journals[2][journals[1]==l],journals[3][journals[1]==l])

b<-aggregate(b[2], by=b[1], sum)

m[b[,1],2]<-b[,2]

JacSim [k,l]<-1-(sum(abs(m[,1]-m[,2])))/(sum(m[,1],m[,2]))

JacSim [l,k]<-1-(sum(abs(m[,1]-m[,2])))/(sum(m[,1],m[,2]))

m[,2]<-0

}

}

}

#Make sure that matrix JacSim only contains positive values.

JacSim<-abs(JacSim)

#Write matrix JacSim as .csv file.

write.csv(JacSim, "D:\\DATA\\Core collectie\\Jacsimilariteiten cited
journals.csv")
```

## Appendix 2: R-script for assigning similarities to journals from which articles are included in the references of WUR articles

Below, the R-script is given that has been used for assigning a similarity to each journal that is included in the reference list of a WUR article. The similarities have been calculated with the R-script provided in appendix 1. Commentary lines in this script start with “#”.

```
#Import datafile articles.csv with five columns. The first column consists #of the publication year of the articles. The second column consists of ISI #number of the articles. The third column consists of numbers for the #journals in which the articles have been published. The fourth column #consists of numbers for the journals to which the citations have been #made. The fifth column is empty but will be filled via this script with #the similarity values that have been calculated between the journals based #upon co-citing these journals in WUR articles published in 2006-2015. The #similarities are obtained from a similarity matrix JacSim. This matrix is #calculated via the script provided in appendix 1. Import datafile #articles.csv.

articles<-read.csv("D:\\DATA\\articles.csv",header=T,sep=",")

#Import similarity matrix JacSim.csv

JacSim<-read.csv("D:\\DATA\\JacSim.csv, header=T, sep=",")

#Loop over all citations in the matrix articles. The number for the journal #in which the article is published will be placed in the variable pub. The #number for the journal to which a citation is made will be placed in the #variable cit. For this combination of journals the similarity will be #obtained from the similarity matrix JacSim and will be written in the #fourth column in the matrix articles.

for (k in 1:nrow(articles)) {

pub<-articles[k,3]

cit<-articles[k,4]

articles[k,5]<-JacSim[pub,cit]

print(k)

}

#Write matrix articles as .csv file.

write.csv(articles, "D:\\DATA\\articles with similarities.csv")
```